\documentclass{iopart}
\usepackage{graphicx}

\expandafter\let\csname equation*\endcsname\relax
\expandafter\let\csname endequation*\endcsname\relax

\usepackage{amsmath,amssymb,indentfirst}
\usepackage{psfrag}
\usepackage{epsfig}
\usepackage[pdftex]{color}
\usepackage{color}
\usepackage[tight]{subfigure}
\definecolor{dred}{rgb}{0.7,0.0,0.0}
\usepackage{dcolumn}
\usepackage{bm,bbm}
\allowdisplaybreaks 
\usepackage{braket}
\usepackage{ulem} 
\usepackage{hyperref}
\hypersetup{colorlinks=true, linkcolor=blue, citecolor=blue, filecolor=blue, urlcolor=blue, pdftitle=, pdfauthor=, pdfsubject=, pdfkeywords=}

\newcommand{\s}[1]{{\mathsf{#1}}} 

\definecolor{orange}{rgb}{1,0.5,0}
\definecolor{black}{rgb}{0,0,0}

\begin{document}

\title{Fractional (Chern and topological) insulators}

\vskip 10pt
\author{
Titus Neupert,$^{1}$
Claudio Chamon,$^{2}$ 
Thomas Iadecola,$^{2}$
Luiz~H.~Santos,$^{3}$
Christopher Mudry$^{4}$
}
\address{
$^1$ Princeton Center for Theoretical
  Science, Princeton University, Princeton, New Jersey 08544, USA
\\
$^2$ Physics Department, Boston University, 
590 Commonwealth Avenue, Boston, MA 02215, USA
\\
$^3$ Perimeter Institute for Theoretical Physics, Waterloo, ON, N2L 2Y5, Canada
\\
$^4$ Condensed Matter Theory Group, Paul Scherrer Institute, 
CH-5232 Villigen PSI, Switzerland
}

\date{\today}

\begin{abstract}
We review various features of interacting Abelian topological phases
of matter in two spatial dimensions, placing particular emphasis on fractional
Chern insulators (FCIs) and fractional topological insulators (FTIs).
We highlight aspects of these systems that challenge the intuition
developed from quantum Hall physics---for instance, FCIs are stable in
the limit where the interaction energy scale is much larger than the
band gap, and FTIs can possess fractionalized excitations in the bulk
despite the absence of gapless edge modes.
\end{abstract}

\maketitle



\section{Introduction}

The theoretical exploration of topological phases of matter is widely based on
the study of model Hamiltonians which in many cases are exactly soluble. 
The analysis of the universal properties of the AKLT model \cite{AKLT}, the toric code \cite{toric}, and string-net models \cite{string-net} in general, has essentially shaped our thinking about topological phases in any dimension. 
However, finding experimental realizations of these phases in electronic quantum matter has proven a hard task in most cases. In that light, it can only be seen as a blessing of nature that the archetypal topological phase of electrons, namely the quantum Hall effect, comes with an exactly soluble Hamiltonian of degenerate Landau levels in the integer case, and, in many instances, with ground states close to model states in the fractional case. 

This blessing, however, came at the price that it obscured the distinction between properties specific to the 
Landau level Hamiltonian on one hand, and universal features of the topological phase on the other hand.
Over the years, several studies introduced deviations from the model Hamiltonian, by considering a periodic background potential \cite{Read93}, an underlying lattice \cite{Kapit10}, and deviations from rotational symmetry \cite{Haldane12}.
Correcting the common beliefs that had emerged from the study of the Landau level problem, they revealed that the Landau level filling fraction need not be the same as the Hall conductivity, and that the zeros in the Laughlin wave function are not tied to the electrons' positions. 

Recent years have witnessed a growing interest in the study of models that depart even further from the Landau level problem while still exhibiting a fractional quantum Hall effect (FQHE). Such models, known as fractional Chern insulators (FCIs), evolved from the insight that the integer quantum Hall effect (IQHE) can be realized in certain lattice band insulators, known as Chern insulators, without an external magnetic field \cite{Haldane88}. Indeed, it was found that FCIs, and the associated FQHE physics, can emerge from the partially filled bands of a Chern insulator upon the introduction of sufficiently strong repulsive interactions between electrons \cite{Tang11,Sun11,Neupert11,Sheng,RB1}.

Beyond a deeper understanding of the quantum Hall effect, Chern insulators and their fractional counterparts have also provided the foundation for an entirely new class of topological phases, distinguished by the presence of time-reversal symmetry (TRS).  The exploration of these phases originated with the theoretical proposal of the quantum spin Hall effect \cite{Kane05a,Kane05b,Bernevig06a,Bernevig06b}, the simplest models of which can be understood as direct products of time-reversed copies of a Chern insulator.  This picture was later generalized to the $\mathbb Z_2$ topological band insulators \cite{Fu-Kane06,Fu07,Moore07,Roy09,Kane-Hasan/Qi-Zhang}, which have been observed experimentally in two \cite{Konig07} and three \cite{Hsieh08,Chen09,Hsieh09c} dimensions.   In parallel to the development of FCIs from Chern insulators, one can begin to explore ways of combining FCIs to build so-called fractional topological insulators (FTIs) \cite{Bernevig06a,LevinStern1,NeupertFTI,SantosFTI,LevinStern2,RepellinFTI}.

In this paper, we provide a perspective on these fractionalized phases of matter.  We begin by reviewing various parallels between FCIs and the FQHE, before highlighting several aspects of FCIs that go markedly beyond the Landau level paradigm.  (For many details about the physics of FCIs, we refer the reader to existing reviews \cite{ParameswaranFCIReview,BergholtzFCIReview}.)  We move on to an overview of FTIs, focusing on features that have no analog in noninteracting topological band insulators.  First, we present an interacting fermionic lattice model that demonstrates how to construct an FTI from two FCIs, and provide a summary of the various phases of this model that can be obtained for different interaction parameters.  Next, we review the Chern-Simons topological field theories for Abelian FTIs, from which their bulk-edge correspondence can be derived.  While they are intrinsically topological in that they support fractionalized quasiparticle excitations in the bulk, FTIs may or may not feature protected gapless edge modes, in contrast with $\mathbb Z_2$ topological insulators, whose primary interest derives from these edge modes.  We review the stability criterion that determines the presence or absence of these edge states in FTIs, and point to several interesting aspects of FTIs with gapped edges that merit further study.

\section{Fractional Chern insulators}

Fractional Chern insulators are two-dimensional lattice systems of interacting, itinerant fermions or bosons whose gapped many-body ground state shares its universal (topological) features with the fractional quantum Hall states that appear in Landau levels. In particular, FCI ground states support a quantized fractional Hall conductivity, universal quasiparticle excitations, a topological ground state degeneracy that depends on the genus of the underlying manifold, as well as universal structure in their entanglement spectrum. 

With an increasing number of numerical studies, it became clear that FCIs emerge in a variety of different microscopic Hamiltonians, in particular as far as the noninteracting part of the models is concerned. Consider a Hamiltonian
\begin{equation}
H_{\mathrm{FCI}}=H_0+H_{\mathrm{int}},
\end{equation} 
where $H_{\mathrm{int}}$ encodes repulsive, short-range, density-density interactions between the bosons or fermions.
For this discussion, we want to focus on the case where the noninteracting Hamiltonian $H_0$ is translationally invariant and supports a spectrally isolated band, that is, the minimal spectral gap between any state in this band and any other single-particle eigenstate of $H_0$ is much larger than the bandwidth of this particular band. It is then meaningful to project the dynamics of the system onto the degrees of freedom of this isolated band. The isolated band is fully characterized by its energy dispersion $\varepsilon_{\boldsymbol{k}}$, and the projector $P_{\boldsymbol{k}}$ onto its Bloch eigenstates for every $\boldsymbol{k}$ in the Brillouin zone (BZ). However, $P_{\boldsymbol{k}}$ is not a gauge-invariant quantity. Instead the relation
\begin{equation}
\mathrm{tr}[P_{\boldsymbol{k}}(\partial_{k_\mu}P_{\boldsymbol{k}})(\partial_{k_\nu}P_{\boldsymbol{k}})]
=: g_{\mu\nu;\boldsymbol{k}}+\frac{\mathrm{i}}{2}F_{\mu\nu;\boldsymbol{k}}
\end{equation}
defines two gauge-invariant tensors: the symmetric real quantum metric $g_{\mu\nu;\boldsymbol{k}}$ and the antisymmetric real Berry connection $F_{\mu\nu;\boldsymbol{k}}$ \cite{Haldane11,Roy12}. Together these quantities characterize the ``quantum geometry'' defined by the band under consideration. The quantum metric defines a distance measure in wavefunction space, while the Berry curvature directly enters the semiclassical equations of motion. The topological significance of $F_{\mu\nu;\boldsymbol{k}}$ is that it integrates to the Chern number
\begin{equation}
C=2\pi\int\frac{\mathrm{d}^2\boldsymbol{k}}{\Omega_{\mathrm{BZ}}}F_{12;\boldsymbol{k}},
\end{equation}
and thus determines the Hall conductivity $\sigma^{\mathrm{H}}_{12}=C\,e^2/h$ of the filled band. (One should note that the $g_{\mu\nu;\boldsymbol{k}}$ and $F_{\mu\nu;\boldsymbol{k}}$ also depend on the real-space embedding of the microscopic degrees of freedom that enter the model \cite{Yang-Le1}.)

We may now define the \emph{Landau level limit} by
\begin{equation}
g_{\mu\nu;\boldsymbol{k}}=\mathrm{const},\quad
F_{\mu\nu;\boldsymbol{k}}=\mathrm{const},\quad
\varepsilon_{\boldsymbol{k}}=\mathrm{const},\quad
|C|=1.
\end{equation}
A band that obeys these properties (to some accuracy) is essentially indistinguishable from a Landau level, despite the fact that it might be defined within a lattice model. Naturally, the low-lying bands of Hofstadter models with sufficiently small flux per plaquette are very close to the Landau level limit. As a consequence, such models also exhibit the FQHE when bands close to the Landau level limit are fractionally filled with interacting particles. 
Models that are in the Landau level limit have a larger translational symmetry than generic lattice models.
Translations on the lattice are implemented by the Fourier component of the density operator $\rho_{\boldsymbol{q}}$, projected to the given band, if $\boldsymbol{q}$ is taken to be the minimal spacing in momentum space. In the Landau level limit, the operators $\rho_{\boldsymbol{q}}$ satisfy a closed algebra \cite{Roy12}
\begin{equation}
[\rho_{\boldsymbol{q}},\rho_{\boldsymbol{q}'}]=2\mathrm{i}\sin\left(F_{12}\frac{\boldsymbol{q}\wedge\boldsymbol{q}'}{2}\right)
e^{q_\mu g_{\mu\nu} q^\prime_\nu}\rho_{\boldsymbol{q}+\boldsymbol{q}'},
\end{equation}
which is known as the Girvin-MacDonald-Platzman algebra \cite{GMP} in the context of the FQHE. Many universal characteristics of the FQHE are attributed to this algebra. 

The remarkable feature of FCIs is that they exhibit FQHE-like behavior even when the band structure deviates significantly from the Landau level limit and lattice effects play a nonperturbative role. It has been shown that any of the properties that define the Landau level limit can be strongly violated in FCIs. 

First, many models that support stable FCI phases have $g_{\mu\nu;\boldsymbol{k}}=0$ or $F_{\mu\nu;\boldsymbol{k}}=0$ for some values of $\boldsymbol{k}$, implying strong fluctuations in the quantum metric and Berry curvature.

Second, FCIs prevail even if the target band is not flat and energetically isolated, but instead other bands (with like or different Chern numbers) are close in energy on the scale of the interaction energy \cite{Kourtis1,Grushin14}. The picture that emerges is that FCIs are favored if the ratio of the interaction energy scale to the bandwidth of the target band is large, irrespective of the locations and properties of other nondegenerate bands. In particular, FCIs can survive the limit of infinitely strong interactions. 

Third, FCIs can emerge in bands with Chern numbers $|C|>1$, in which case a rich hierarchy of states at filling factors
\begin{equation}
\nu=\frac{k}{2|C|+1}, \qquad \nu=\frac{k}{|C|+1},\quad k\in\mathbb{Z},
\end{equation}
was discovered for fermions and bosons, respectively \cite{Sheng12,Grushin12,Lauchli,Bernevig12}. These states have no analogues in an isolated, nondegenerate Landau level, which is bound to have Chern number 1. One might wonder whether multi-layer FQH states, which effectively have a higher Chern number, can support equivalent states. However, the $|C|>1$ FCIs differ from such states in their boundary conditions, which entail modified exclusion statistics \cite{Yang-Le2}. Aside from bands with $|C|\geq 1$, FCIs can also arise in band structures with $C=0$, if, for example, the model is time-reversal-invariant. It is then necessary that the interaction break the TRS spontaneously to form an FCI. (We will discuss an example for this scenario in the next section.)
More generally, FCIs can coexist with a variety of symmetry-broken orders, such as charge density waves \cite{Kourtis2}, and the presence of a lattice provides a rich pool of possible competing phases.

To summarize this section, we characterized FCIs as a class of systems that exhibit FQH-like topologically ordered ground states and generalize the Landau level Hamiltonians of the FQHE. The Landau level Hamiltonian is a special  
 case of an FCI, in which the energy dispersion, the quantum metric, and the Berry curvature are independent of momentum. Hence, FCIs can mimic even to high quantitative precision the FQH states in Landau levels (including low-lying excited states \cite{Repellin14}), but can also go markedly beyond them in many ways.
 
So far, FCIs remain a theoretical possibility without experimental realization. However, we would like to point to several encouraging experimental developments that have taken place in recent years.
By now, several instances of a Chern insulator, the noninteracting system that underlies the FCI, have been demonstrated in a variety of platforms.  ``Classical'' realizations of Chern insulators in photonic wave guides are not relevant for building FCIs, because they cannot provide access to interacting quantum ground states. Much more appropriate are the realizations of a Hofstadter model \cite{Bloch} and the Haldane model \cite{Esslinger} in optical lattices filled with ultracold atoms. Finally, Chern insulators have also been demonstrated in an electronic condensed matter system, by magnetically doping a thin-film topological insulator.
Of these approaches, ultracold atomic gases seem to offer the best prospects for realizing FCIs. They allow for significant tunability of the interactions via Feshbach resonances and work for fermions and bosons alike, with the latter expected to exhibit more stable FCI states. However, ultracold atomic gases come with their own unique challenges, such as state preparation \cite{Laumann}, the problem of experimental verification of an FCI state, and various consequences of their driven, out-of-equilibrium nature \cite{GrushinFFCI}.

\section{Fractional topological insulators}

Having surveyed our understanding of FCIs, we now turn to the case
where TRS is present. The study of
this class of systems is inspired by the discovery of the
$\mathbb{Z}_2$ topological band insulators. Indeed, in light of this discovery, it is natural to ask what are the possible
fractionalized phases descending from the $\mathbb{Z}_2$ topological
band insulators. Below, we review results that start to answer this
question.

A simple example of a fractional fluid with TRS can be obtained by putting
together two copies of a FQH system with opposite chiralities for up
and down spins \cite{Bernevig06a,Freedman04,Hansson04}. Such systems
display a fractional quantum spin Hall effect (FQSHE). Levin and Stern \cite{LevinStern1}
analyzed the stability of the edge states of these doubled systems
against interactions and disorder that respect TRS.  Bosonic analogues of these FTIs
have also been studied, and were found to be robust to perturbations in the bulk that couple
the two spin species at both the single-particle and many-body levels \cite{RepellinFTI}.

Indeed, FTIs can be defined in greater generality than the above description in terms of direct products of time-reversed FQH systems. Our discussion in this section is motivated by several related observations.  
First, there is the practical question of how to realize two copies of a FQH system with opposite signs of
the fractional Hall conductance. For example, it is difficult to devise a situation in which large magnetic fields pierce two parallel planes in opposite directions.  (It may be simpler to realize such a system in electron- and
hole-doped bilayers instead.) We propose that one possible realization
of an FTI is in lattice models for $\mathbb{Z}_2$ topological band insulators
with flat topological bands and appropriate interactions. Second, we would like
a unified description of Abelian FTIs that are more general than just two copies of the
FQHE. To that end, we consider below the most generic Abelian fractional states
compatible with TRS. (We note that our perspective is still limited in
that we consider only Abelian states; for studies of non-Abelian
states, see Ref.~\cite{Cappelli}.) Third, we would like to stress
that the existence of gapless edge states is too restrictive a condition
to select which TRS topological fluids should be deemed non-trivial. As
opposed to non-interacting $\mathbb{Z}_2$ topological band insulators,
which, apart from gapless edge states, have no physical telltale of a
non-trivial nature, FTIs with gapped edges can still display
non-trivial bulk properties, such as quasiparticles with fractional charge and statistics.  Such fractionalized liquids host a topological degeneracy on closed surfaces of
non-zero genus, which is a defining property of topologically ordered states.
Notably, FTIs with gapped edges can also host a topological degeneracy on
planar surfaces with holes, which are in principle easier to probe experimentally.
Thus, FTIs are candidates to provide experimental evidence for topological order.

We elaborate on these three points below.  We first present a lattice model for an FTI and discuss its phase diagram, before moving on to discuss the time-reversal-symmetric Abelian Chern-Simons theories that encode the topological features of FTIs.  We review the stability criterion for gapless edge modes before discussing the topological degeneracy of FTIs with gapped edges on planar manifolds with holes.

\subsection{Lattice realization of an FTI}

Consider the Hamiltonian
\begin{equation}
H=H^{\ }_0+H^{\ }_{\mathrm{int}}
\;,
\label{eq:H lattice}
\end{equation}
broken down into a quadratic part $H^{\ }_0$ and an interacting one
$H^{\ }_{\mathrm{int}}$. The piece $H^{\ }_0$ we choose to be made of
two copies of the $\pi$-flux phase with flat bands that was studied in
Ref.~\cite{Neupert11}, one copy for each of the $\uparrow$ and
$\downarrow$ spin species. The model is defined on the square lattice,
whose two sublattices we denote by $A$ and $B$. We define the 
spinor
$\psi^\dagger_{\boldsymbol{k},\sigma}:= \left(
  c^\dagger_{\boldsymbol{k},\mathrm{A},\sigma},
  c^\dagger_{\boldsymbol{k},\mathrm{B},\sigma} \right)$,
where
$c^\dagger_{\boldsymbol{k},\mathrm{A},\sigma}$
creates an electron with spin $\sigma=\uparrow,\downarrow$ and Bloch momentum $\boldsymbol{k}$
in the reduced Brillouin zone of sublattice A.
We then write
\begin{subequations}
\begin{equation}
H^{\ }_0:=
\sum_{\boldsymbol{k}\in \mathrm{BZ}}
\left(
\psi^\dagger_{\boldsymbol{k},\uparrow}
\;
\widehat{\boldsymbol{B}}^{\ }_{\boldsymbol{k}}\cdot\boldsymbol{\tau}
\;
\psi^{\ }_{\boldsymbol{k},\uparrow}
+
\psi^\dagger_{\boldsymbol{k},\downarrow}
\;
\widehat{\boldsymbol{B}}^{\ }_{-\boldsymbol{k}}\cdot\boldsymbol{\tau}^{\mathsf{T}}
\;
\psi^{\ }_{\boldsymbol{k},\downarrow}
\right),
\label{eq: lattice H0}
\end{equation}
where $\widehat{\boldsymbol{B}}^{\ }_{\boldsymbol{k}}\equiv {\boldsymbol{B}}^{\
}_{\boldsymbol{k}}/\left|\boldsymbol{B}^{\ }_{\boldsymbol{k}}\right|$ has components
\begin{eqnarray}
&&
B^{\ }_{1,\boldsymbol{k}}
+
\text{i}
B^{\ }_{2,\boldsymbol{k}}
:=
t^{\ }_{1}\,e^{-\text{i}\pi/4}
\left(
1
+
e^{
+\text{i}\left(k^{\ }_{y}-k^{\ }_{x}\right)
  }
\right)
\nonumber\\
&&
\hphantom{
B^{\ }_{1,\boldsymbol{k}}
+
\text{i}
B^{\ }_{2,\boldsymbol{k}}
:=
         }
+
t^{\ }_{1}\,e^{+\text{i}\pi/4}
\left(
e^{
-\text{i}k^{\ }_{x}
  }
+
e^{
+\text{i}k^{\ }_{y}
  }
\right),
\\
&&
B^{\ }_{3,\boldsymbol{k}}
:=
2t^{\ }_{2}
\left(\cos k^{\ }_{x}-\cos k^{\ }_{y}\right).
\end{eqnarray}
\end{subequations}
The three Pauli-matrices $\boldsymbol{\tau}=(\tau^{\ }_1,\tau^{\ }_2, \tau^{\
}_3)$ act on the sublattice index.  Here, $t^{\ }_1$ and $t^{\ }_{2}$
represent the nearest neighbor (NN) and next-nearest neighbor (NNN)
hopping amplitudes. The model has flat bands (because of the
normalization $\left|\boldsymbol{B}^{\ }_{\boldsymbol{k}}\right|$), and it is local
(in the sense of exponentially decaying hoppings in position space) as
long as $t_2\ne 0$, which ensures that $\left|\boldsymbol{B}^{\
  }_{\boldsymbol{k}}\right|$ never vanishes.

One verifies that $H^{\ }_{0}$ is invariant under TRS, and that the
$z$-component of spin is conserved (since the particle numbers of the two spin species are conserved separately). The lower bands for the $\uparrow,\downarrow$ spins
have opposite Chern numbers $\pm 1$, and hence if the system were
half-filled, it would have spin-Hall conductivity
$\sigma^{\mathrm{SH}}_{\mathrm{xy}}=2\times e/(4\pi)$.

The Hamiltonian $H^{\ }_{0}$ consists of two copies of a flat-band Chern insulator. 
Like the Kane-Mele model \cite{Kane05a}, it arises due to spin-orbit interactions.
Although the model effectively corresponds to two time-reversed copies of a system with quantized
Hall response, there is no magnetic field
required. In this sense, the Hamiltonian $H^{\ }_{0}$ provides a microscopic route to obtaining an
FTI, should a partially-filled version of the model sustain liquid
ground states when interactions are switched on. We thus turn to the
interactions to be added.

Consider repulsive interactions defined in terms of the spin densities
$\rho^{\ }_{i,\sigma}$ at site $i$ and spin $\sigma$:
\begin{equation}
\begin{split}
H^{\ }_{\mathrm{int}}
:=&\,
U
\sum_{i\in \Lambda}
\rho^{\ }_{i,\uparrow}
\rho^{\ }_{i,\downarrow}
+
V
\sum_{\left\langle ij\right\rangle\in \Lambda}
\Big[
\rho^{\ }_{i,\uparrow}
\rho^{\ }_{j,\uparrow}
+
\rho^{\ }_{i,\downarrow}
\rho^{\ }_{j,\downarrow}
\\
&\,+
\lambda
\left(
\rho^{\ }_{i,\uparrow}
\rho^{\ }_{j,\downarrow}
+
\rho^{\ }_{i,\downarrow}
\rho^{\ }_{j,\uparrow}
\right)
\Big],
\end{split}
\end{equation}
where $\left\langle ij\right\rangle$ are directed NN bonds of the
square lattice $\Lambda=\mathrm{A}\cup\mathrm{B}$. We consider both
onsite Hubbard interactions $U$ and nearest-neighbor interactions
$V$. The parameter $\lambda$ defines the spin anisotropy; the
interaction part of the model is SU(2)-symmetric when
$\lambda=1$. When $U=0$ and $\lambda=0$, the model corresponds to two
copies of the FCI considered in Ref.~\cite{Neupert11} with
opposite fractional quantum Hall conductivities. Notice that $H^{\
}_{\mathrm{int}}$, like $H^{\ }_0$, does not break the separate
$\uparrow,\downarrow$ number conservation.

The nature of the ground state of the interacting Hamiltonian $H=H^{\
}_0+H^{\ }_{\mathrm{int}}$ depends on the filling of the
$\mathbb{Z}_2$ topological bands, and on the details of the
interactions. Exact diagonalization, with projection into the lower band, was employed in
Ref.~\cite{NeupertFTI}. The
following phases were found in different regimes, when the lattice is
$1/3$ filled (or, equivalently, when $2/3$ of the lower band is filled).

\subsubsection{FTI phase}\label{sec: FTI phase}

If $U=0$ and $\lambda=0$, the problem is equivalent to two decoupled
FCIs, with opposite chiralities. This is a lattice realization of two
FQHE layers with opposite Hall conductance. We identify the state
numerically by computing the spin Hall conductance, and by analyzing
the ground state degeneracy on the torus and the evolution of the ground-state
manifold with inserted fluxes (coupling independently to the up and
down spin components). The ground state manifold is 9-fold degenerate,
which is to be expected from independent layers with 3-fold degeneracy
each.

\subsubsection{Spontaneously magnetized FCI phase}

For large onsite interaction $U$ (say $U/V=3$) and isotropic NN interactions ($\lambda=1$) we find that TRS is spontaneously
broken, and the ground state is ferromagnetic. The ground state is in
the sector where the occupation numbers for one of the two spin
species is maximal in the lower band, and the other species fills
$1/3$ of the available states ($2/3+2/3=1+1/3$). The ground state, in
addition to the two-fold degeneracy due to its ferromagnetic order,
has a three-fold degeneracy associated with an FCI with filling
$\nu=1/3$. Therefore the total ground state degeneracy is 6. This is
an example of the coexistence of symmetry-broken order and topological
order (c.f.~Ref.~\cite{Kourtis2}).

\subsubsection{Mysterious phase}

In addition to the two phases that we can identify above, we have
found an unknown phase that appears when the nearest neighbor interaction is isotropic
($\lambda=1$) and the onsite interaction $U$ is negligible as compared
to the NN interaction strength $V$ (say $U=0$). This state is not
ferromagnetic; the ground state has equal populations of up and down
particles.  Furthermore, we have no evidence that it is a charge density wave. 
It is, however, three-fold degenerate. As we shall see
below, when we discuss a general description of TRS Abelian
Chern-Simons theories, the degeneracy of the ground state is the square of
an integer. Since the ground state degeneracy that we find in this
example is 3, this system is not in such an Abelian FTI state; it
might possibly be an interesting non-Abelian FTI.  A subsequent exact diagonalization
study of the Hamiltonian \eqref{eq:H lattice} explored the quantum phase transition
between the FTI phase at $(U,\lambda)=(0,0)$, mentioned in Sec.~\ref{sec: FTI phase} above, and this mysterious phase at $(U,\lambda)=(0,1)$ \cite{Sheng14}.
The onset of the transition is signalled by the closing of the energy and quasispin excitation gaps.
Despite this phase transition, which lifts the 9-fold topological ground-state degeneracy to a 3-fold one,
that study found that the spin Chern number \cite{Sheng06} is the same on both sides of the transition.  
Elucidating the nature of this state demands further work.

\subsection{Time-reversal-symmetric Abelian Chern-Simons
theories}

Below we present a general description of FTIs using multi-component
Abelian Chern-Simons fields. This description generalizes the
hierarchy of Abelian FQHEs to the hierarchy of Abelian fractional
quantum spin Hall effects (FQSHEs). The details of this bulk
construction were originally presented in
Ref.~\cite{SantosFTI}, and the analysis of the stability of
the edge states in Ref.~\cite{NeupertFTI}.

The FTIs are described by a time-reversal symmetric (2+1)-dimensional
Chern-Simons quantum field theory that depends on $2N$ flavors of
gauge fields $a^{\ }_{i,\mu}(t,\boldsymbol{x})$, where $i=1,\dots,2N$ labels
the flavors and $\mu=0,1,2$ labels the space-time coordinates
$x^{\mu}\equiv(t,\boldsymbol{x})$, and with the action
\label{eq:intro-CS}
\begin{equation}
\begin{split}
\mathcal{S}:=
\int\mathrm{d}t\,\mathrm{d}^{2}\boldsymbol{x}\;
&
\epsilon_{\ }^{\mu\nu\rho} \,
\left(
-
\frac{1}{4\pi}
K^{\ }_{ij}\;
a^{\ }_{i,\mu}\;
\partial^{\ }_{\nu}\,
a^{\ }_{j,\rho}
\right.
\\&
\left.
+
\frac{e}{2\pi}\,Q^{\ }_{i}\,
A^{\ }_{\mu}\,
\partial^{\ }_{\nu}\,
a^{\ }_{i,\rho}
+
\frac{s}{2\pi}\,S^{\ }_{i}\,
B^{\ }_{\mu}\,
\partial^{\ }_{\nu}\,
a^{\ }_{i,\rho}
\right).
\end{split}
\label{eq:intro-CS a}
\end{equation}
Here, $K^{\ }_{ij}$ are elements of the symmetric and invertible
$2N\times2N$ integer matrix $K$. The integer-valued component
$Q^{\,}_{i}$ of the $2N$-dimensional vector $Q$ represents the $i$-th
electric charge in units of the electronic charge $e$, which couples
to the electromagnetic gauge potential $A_{\mu}(t,\boldsymbol{x})$. Similarly,
$S^{\,}_{i}$ is an integer-valued component of the $2N$-dimensional
vector $S$ that represents the $i$-th spin charge in units of $s$
associated to the up or down spin projection along a spin-1/2
quantization axis, which couples to the Abelian (spin) gauge potential
$B^{\,}_{\mu}(t,\boldsymbol{x})$. There is a compatibility condition for the
quasiparticles to obey consistent ststistics, namely that $(-1)^{K^{\ }_{ii}}=
(-1)^{Q^{\ }_{i}},i=1,\cdots,2N$.

The operation $\mathcal{T}$ for reversal of time
has the following action on the CS fields
\begin{eqnarray}
a^{\ }_{{i},\mu}(t,\boldsymbol{x})
\stackrel{\mathcal{T}}{\rightarrow}
\begin{cases}
-g^{\mu\nu}\,
a^{\ }_{{i}+N,\nu}(-t,\boldsymbol{x})
&\mbox{if } {i}=1,\cdots,N, 
\\
-g^{\mu\nu}\,
a^{\ }_{{i}-N,\nu}(-t,\boldsymbol{x})
&\mbox{if } {i}=N+1,\cdots,2N,
\end{cases} 
\nonumber
\end{eqnarray}
for $\mu=0,1,2$, where $g^{\ }_{\mu\nu}:=\mathrm{diag}(+,-,-)\equiv
g^{\mu\nu}$ is the Lorentz metric.  

TRS imposes that the matrix $K$ and the vectors $Q$ and $S$ are of the
block form
\begin{equation}
K=\left(
\begin{matrix}
    \kappa&\;\Delta\\
    \Delta^{\!\mathsf{T}}&-\kappa
\end{matrix}
\right),
\quad
Q=\left(
\begin{matrix}
\varrho\\
\varrho
\end{matrix}
\right),
\quad
S=\left(
\begin{matrix}
\,\varrho\\
-\varrho
\end{matrix}
\right),
\label{eq:intro-K-matrix}
\end{equation}
with $\varrho$ an integer $N$-vector, while
$\kappa=\kappa^{\!\mathsf{T}}$ and $\Delta=-\Delta^{\!\mathsf{T}}$ are
symmetric and antisymmetric integer-valued $N\times N$ matrices,
respectively. The special cases of the FQSHE treated in
Refs.~\cite{Bernevig06a} and \cite{LevinStern1} correspond to
setting $\Delta=0$ in the $K$ matrix above. 

Notice that if one computes the charge Hall
conductivity~\cite{WenZee92}, one obtains
\begin{equation}
\sigma_{H}=\frac{e^2}{h}\;
Q^{\mathsf{T}}\,K^{-1}\,Q
=0,
\label{eq: def nu a}
\end{equation}
which is expected for systems respecting TRS. One can therefore think of the
states represented by the $K$ matrix and $Q$ and $S$ vectors above as
``$\nu=0$'' FQH states, owing to the 
doubled structure of the theory.

This doubled structure is even more evident if we express
the CS theory as a BF theory~\cite{Blau91,Hansson04,SantosFTI}, i.e., 
by defining 
\begin{subequations}
\begin{equation}
a^{(\pm)}_{\s{i},\mu}
:=
\frac{1}{2}
\left(
a^{\ }_{\s{i},\mu}
\pm
a^{\ }_{\s{i}+N,\mu}
\right),
\qquad \s{i}=1,\dots,N
\;,
\label{eq: def a +- a}
\end{equation}
for $\mu=0,1,2$.
This basis allows one to re-express the effective action%
~(\ref{eq:intro-CS a}) as
\begin{equation}
\mathcal{S}:=
\!\!\int\!\! \,\mathrm{d}t\, \mathrm{d}^{2}\boldsymbol{x}\; 
\epsilon^{\mu\nu\rho} 
\left(
-
\frac{1}{\pi}
\varkappa^{\ }_{\s{ij}}\,
a^{(+)}_{\s{i},\mu}\,
\partial^{\ }_{\nu}\,
a^{(-)}_{\s{j},\rho}
+ 
\frac{e}{\pi}\;
\rho^{\ }_{\s{i}}\,
A^{\ }_{\mu}\partial^{\ }_{\nu}\,
a^{(+)}_{\s{i},\rho}
+ 
\frac{s}{\pi}\;
\rho^{\ }_{\s{i}}\,
B^{\ }_{\mu}\partial^{\ }_{\nu}\,
a^{(-)}_{\s{i},\rho}
\right).
\label{eq:intro-CS-BF}
\end{equation}
In this representation, the indices $\s{i},\s{j}$ run from $1$ to $N$
(as opposed to $i,j$ which run from $1$ to $2N$), while the coupling
between the pair of gauge fields $a^{(+)}$ and $a^{(-)}$ is
off-diagonal in the BF labels $(\pm)$. This is a consequence of
time-reversal symmetry, implemented by
\begin{equation}
a^{(\pm)}_{\mu}({t,\boldsymbol{x}})
\stackrel{\mathcal{T}}{\rightarrow}
\mp g^{\mu\nu}\, 
a^{(\pm)}_{\nu}({-t,\boldsymbol{x}})
\;,
\label{eq: def TR bf}
\end{equation}
which leaves the Lagrangian density%
~(\ref{eq:intro-CS-BF}) invariant. In this representation, the 
electromagnetic gauge potential $A$ couples to the
$(+)$-species only, while the spin gauge potential $B$ couples to the 
$(-)$-species only. The $N\times N$ integer-valued matrix $\varkappa$ 
in the BF representation is related to the 
block matrices $\kappa$ and $\Delta$ contained in $K$ from 
Eq.~(\ref{eq:intro-K-matrix})
through
\begin{equation}
\varkappa=
\kappa
-
\Delta.
\end{equation}
\end{subequations}

The topological ground state degeneracy on the torus is obtained for either
description, 
i.e., the one in terms of the flavors 
$a^{\ }_{i}$ with $i=1,\cdots,2N$ 
or the one in terms of the flavors 
$a^{(\pm)}_{\s{i}}$ with $\s{i}=1,\cdots,N$, 
from \cite{WenZee92,Wesolowski94}
\begin{eqnarray}
\label{eq:intro-degeneracya}
\mathcal{N}^{\rm torus}_{\mathrm{GS}}&=&
\left|
\mathrm{det}
\begin{pmatrix}
\kappa
&
\Delta
\\
\Delta^{\!\mathsf{T}}
&
-\kappa
\end{pmatrix}
\right|
\nonumber\\
&=&
\left|
\det
\left(
\begin{matrix}
\Delta^{\!\mathsf{T}}
&
-\kappa
\\
\kappa
&
\;\Delta
\end{matrix}
\right)
\right|
\nonumber\\
&=&
\left|
{\rm Pf}
\left(
\begin{matrix}
\Delta^{\!\mathsf{T}}
&
-\kappa\\
\kappa
&
\;\Delta
\end{matrix}
\right)
\right|^2
\nonumber\\
&=&
\left(
\mathrm{integer}
\right)^2,
\end{eqnarray}
or
\begin{eqnarray}
\label{eq:intro-degeneracyb}
\mathcal{N}^{\rm torus}_{\mathrm{GS}}&=&
\left|
\det
\begin{pmatrix}
0
&
\varkappa
\\
\varkappa^{\!\mathsf{T}}
&
0
\end{pmatrix}
\right|
=\left(\det \varkappa\right)^{2}
\nonumber\\
&=&
\left(
\mathrm{same\; integer\;as\;above}
\right)^2.
\end{eqnarray}

\subsubsection{Stability criterion for gapless edge modes}

Whether or not the doubled Chern-Simons theory defined in Eq.~\eqref{eq:intro-CS a} supports stable gapless edge modes in open geometries
can be determined from the $K$ matrix and the integer chrage vector $Q$. The
generic approach presented in Ref.~\cite{NeupertFTI}, which
applies to arbitrary $K$ and $Q$ that respect TRS, was inspired by the
stability analysis of the edge states performed for the single-layer
FQHE by Haldane in Ref.~\cite{Haldane95} (see also Refs.
~\cite{Kane94} and \cite{Moore98}), by Naud et al.\ in
Refs.~\cite{Naud00} and \cite{Naud01} for the bilayer
FQHE, and especially by Levin and Stern in
Ref.~\cite{LevinStern1} for the FQSHE. 
(The cases studied by Levin and Stern correspond to $\Delta=0$ above.)

The stability of the edge states against disorder and interactions is
decided based on whether the integer
\begin{equation}
R:=
r\,
\varrho^{\mathsf{T}}\,
(\kappa-\Delta)^{-1}\,
\varrho
\label{eq: def R}
\end{equation}
is odd (stable) or even (unstable). 
The vector $\varrho$ together with the matrices
$\kappa$ and $\Delta$ were defined in Eq.%
~(\ref{eq:intro-K-matrix}).
The integer $r$ is the smallest integer
such that all the $N$ components of the vector 
$r\,(\kappa-\Delta)^{-1}\, \varrho$ are integers.  Notice that if we restrict to the case when $\kappa=\mathbbm 1_{N}$, $\Delta=0$ and $\varrho^{\mathsf{T}}=(1,1,\dots,1)$ gives
$R=N$, i.e., we have recovered the same criterion as for the
two-dimensional non-interacting $\mathbb{Z}^{\ }_{2}$ topological band
insulator.  Furthermore, if we relax this restriction to require $\Delta=0$ and $\kappa \neq 0$, then the integer $R$ can be interpreted as the ratio of the spin-Hall conductance $\sigma_{\rm xy}^{\rm SH}$ to the smallest quasiparticle charge $e^*$ ($=1/r$) that is allowed in the system, as argued in Ref.~\cite{LevinStern1}.

We would like to contrast the interacting
systems above with the case of the non-interacting
$\mathbb{Z}^{\ }_{2}$ topological insulators. In the latter case, the bulk
is a gapped band insulator that supports no interesting excitations. Therefore, the presence or absence of gapless
edge states, which are the only physical
manifestation of the $\mathbb{Z}^{\ }_{2}$ index, is the defining property of the $\mathbb{Z}^{\ }_{2}$ topological insulators. In contrast, the CS theories above describe gapped systems that can have rich bulk
excitations with fractional charge and fractional mutual
statistics. These FTIs can display a topological degeneracy when the
system is defined in a manifold without boundaries, such as the
torus. If one takes a manifold with boundaries (such as a disk or a
punctured disk), one could still separate these systems into two
classes, those with and without gapless edge modes. Those with gapless
edge modes share a property with both the non-interacting
$\mathbb{Z}^{\ }_{2}$ topological band insulators and the FQHE, in
that they have these gapless edge modes that cannot be localized by
disorder. However, the FTIs without gapless edge modes are notable in that
the topological degeneracy that exists in manifolds without boundaries
also manifests in manifolds {\it with} boundaries. The analysis of
this topological degeneracy is our next topic.

\subsubsection{Topological degeneracy of FTIs with gapped edges in planar geometries}

The ground state degeneracy of interacting topological states of matter defined
on manifolds without boundaries, such as the torus, is a hallmark of
topological order~\cite{Wen89,Wen90,Wen91}. While this degeneracy provides a clear definition of
topological order, and is a useful tool in identifying this type of order
in numerical studies, the topological degeneracy is not simple to
probe experimentally. Among the chief obstacles is the construction of
toroidal samples. In systems like FCIs or the FQHE, once the system
is laid flat and has boundaries, there will be gapless edge
modes. Thus, in these systems, it is not possible to probe a small degenerate ground
state subspace separated by a gap to all other states. However, it is possible to do so in the case of an FTI without gapless
edge states.

Let us focus on Eq.~(\ref{eq:intro-CS}) without the $A^{\ }_{\mu}$ and
$B^{\ }_{\mu}$ gauge potentials, and use two separate sets of $N$ CS
fields $\alpha_{\mu}^{\mathsf{i}}(\boldsymbol{x},t)\equiv
a_{\mu}^{\mathsf{i}}(\boldsymbol{x},t)$ and
$\beta_{\mu}^{\mathsf{i}}(\boldsymbol{x},t)\equiv
a_{\mu}^{{\textsf{i}}+N}(\boldsymbol{x},t)$, for
${\textsf{i}}=1,\cdots,N$. Written in terms
of these fields, the action on the annulus $A$ is given by
\begin{equation}
S^{\,}_{\rm{CS}}=\int\limits\mathrm{d}t
\int\limits_{A}\mathrm{d}^{2}x\, 
\frac{\epsilon^{\mu\nu\rho}}{4\pi}\,
\left[
\kappa^{\,}_{\textsf{ij}}\, \left(
\alpha^{\mathsf{i}}_{\mu}\,\partial^{\,}_{\nu}\,\alpha^{\textsf{j}}_{\rho}
-
\beta^{\mathsf{i}}_{\mu}\,\partial^{\,}_{\nu}\,\beta^{\textsf{j}}_{\rho}
\right)
+ 
\Delta^{\,}_{\textsf{ij}}
\left(
\alpha^{\mathsf{i}}_{\mu}\,\partial^{\,}_{\nu}\,\beta^{\textsf{j}}_{\rho}
-
\beta^{\mathsf{i}}_{\mu}\,\partial^{\,}_{\nu}\,\alpha^{\textsf{j}}_{\rho}
\right)
\right].
\label{annulus}
\end{equation} 
Its transformation law under any local gauge transformation of the
form
\begin{subequations}
\begin{equation}\label{g-trans}
\alpha_{\mu}^{\mathsf{i}} \to \alpha_{\mu}^{\mathsf{i}} +
\partial^{\,}_{\mu}\, \chi_\alpha^{\mathsf{i}}, \qquad
\beta_{\mu}^{\mathsf{i}} \to \beta_{\mu}^{\mathsf{i}} +
\partial^{\,}_{\mu}\, \chi_\beta^{\mathsf{i}},
\end{equation}
where $\chi_\alpha^{\mathsf{i}}$ and $\chi_\beta^{\mathsf{i}}$ are
real-valued scalar fields, is
\begin{equation}
S^{\,}_{\rm{CS}}\to S^{\,}_{\rm{CS}}+\delta S^{\,}_{\rm{CS}}
\end{equation}
with the boundary contribution
\begin{eqnarray}
\delta S^{\,}_{\rm{CS}}\:=\, \int\limits\mathrm{d}t
\oint\limits_{\partial A}\mathrm{d}x^{\,}_{\mu}\,
&&
\frac{\epsilon^{\mu\nu\rho}}{4\pi}\,
\left[
\kappa^{\,}_{\textsf{ij}} \left( \chi_\alpha^{\mathsf{i}}\,
\partial^{\,}_{\nu}\, \alpha_{\rho}^{\textsf{j}} -
\chi_\beta^{\mathsf{i}}\, \partial^{\,}_{\nu}\,
\beta_{\rho}^{\textsf{j}} \right) \right.\nonumber \\ 
&&
\!\!\!
\left.  +
\Delta^{\,}_{\textsf{ij}} \left(
\chi_\alpha^{\mathsf{i}}\,\partial^{\,}_{\nu}\,\beta_{\rho}^{\textsf{j}}
-
\chi_\beta^{\mathsf{i}}\,\partial^{\,}_{\nu}\,\alpha_{\rho}^{\textsf{j}}
\right)
\right].
\label{anomaly}
\end{eqnarray}
\end{subequations}
Here, the boundary $\partial A$ of $A$ is the disjoint union of two
circles ($\partial A\:= S^{1}\sqcup S^{1}$) and $\mathrm
dx_\mu\:=\epsilon_{\mu0\sigma}\, \mathrm d\ell^\sigma$, with $\mathrm
d\ell^\sigma$ the line element along the boundary.

Invariance of the Chern-Simons theory~\eqref{annulus} under gauge transformations of the form~\eqref{g-trans} can be achieved by demanding the existence of gapless edge modes with an action $S^{\,}_{\rm E}$ that transforms as $S^{\,}_{\rm{E}}\to S^{\,}_{\rm{E}}-\delta S^{\,}_{\rm{CS}}$. However, if these edge modes become pinned, then
there are no such excitations at the boundary to compensate for the bulk contribution $\delta S^{\,}_{\rm CS}$. Therefore in a system
with gapped edges, the anomalous term $\delta S^{\,}_{\rm CS}$ must vanish 
identically. This can be accomplished by imposing the following two
``gluing'' conditions for all ${\textsf{i}}=1,\cdots,N$:
\begin{equation}
\chi_\alpha^{\mathsf{i}}\vert^{\,}_{\partial A}=
T^{\,}_{\mathsf{i}\mathsf{j}}\;
\chi_\beta^{\,\mathsf{j}}\vert^{\,}_{\partial A}, \qquad
\alpha^{\mathsf{i}}_{\mu}\vert^{\,}_{\partial A}=
T^{\,}_{\mathsf{i}\mathsf{j}}\;
\beta^{\,\mathsf{j}}_{\mu}\vert^{\,}_{\partial A}
\label{gluing},
\end{equation}
where the $T^{\,}_{\mathsf{i}\mathsf{j}}$ are the elements of an invertible matrix that must satisfy
\begin{equation}
T^\textsf{T}\kappa\, T - \kappa + T^\textsf{T}\Delta-\Delta T=0,
\end{equation}
so that $\delta S_{\rm CS}=0$. These results are discussed in much
detail in~\cite{Iadecola14}. A connection between the
$T^{\,}_{\mathsf{i}\mathsf{j}}$ and the tunneling vectors that gap the
edges is there presented. Using the above conditions, it is possible
to obtain the degeneracy on the annulus in much the same way as one obtains
its counterpart on the torus \cite{Iadecola14}. The degeneracy can be shown to be 

\begin{equation}
\label{eq:degeneracy-annulus}
\mathcal{N}^{\rm annulus }_{\mathrm{GS}}=
\sqrt{\mathcal{N}^{\rm torus }_{\mathrm{GS}}}=|\det\varkappa|
\end{equation}

We stress again that this ground state degeneracy is now attainable in
a planar sample with holes, which is simpler to realize experimentally
than a toroidal geometry. The result can be extended to the case when
there are $n_{\rm h}$ holes in a planar sample, each gapped in the same way. In this case, the
degeneracy is $\sqrt{|{\rm det}\, K|}^{\, n^{\,}_{\rm h}}$. This
degeneracy could, in principle, be detected via heat capacity
measurements, as discussed in~\cite{Iadecola14}.

There exists another class of topologically ordered systems without gapless edge states, namely the surface codes \cite{Bravyi,Freedman01}, which can be understood as constrained quantum spin systems defined on manifolds with boundary.  The topologically degenerate ground states of these systems have been proposed as practical platforms for quantum computation \cite{Fowler,Barends}.  Doubled Chern-Simons theories like those discussed in this section can be thought of as effective low-energy descriptions of these lattice models, in much the same way as the toric code can be described in terms of such a theory \cite{Freedman04,Hansson04}.

\subsection{Conclusion}
In this section, we presented an overview of several results that highlight the distinctive features of FTIs.  We began by presenting numerical
evidence that it is possible to realize time-reversal-symmetric
fractionalized phases in interacting systems whose noninteracting bands possess a nontrivial
$\mathbb{Z}_2$ topological index. We then moved on to review results on the generic
field-theoretic description of Abelian fractional topological liquids with
TRS, and the stability criterion for their gapless edge states. Finally, we discussed a major difference between interacting and
noninteracting time-reversal-symmetric topological states, namely that, in the former case, one
can have nontrivial fractionalized states without gapless edge modes.  These systems can possess a topological degeneracy on planar surfaces with holes, which opens up the possibility that this degeneracy could be measured in experiments, were a suitable system to be discovered.  Such a measurement, if successful, would constitute smoking-gun evidence for topological order.

\medskip

\noindent\textbf{Acknowledgments}

It is a pleasure to thank Adolfo Grushin, Stefanos Kourtis, Nicolas Regnault, Cecile Repellin, and Shinsei Ryu for collaboration on the work discussed here, and on related projects.  We 
acknowledge the Condensed Matter Theory Visitors' Program at Boston University for support.  T.N. was supported by DARPA SPAWARSYSCEN Pacific N66001-11
-1-4110, and C.C. was supported by DOE Grant DEF-06ER46316. T.I. was supported by the National Science Foundation Graduate Research Fellowship Program 
under Grant No. DGE-1247312.  Research at the Perimeter Institute is supported by the Government of Canada through Industry Canada and by the Province 
of Ontario through the Ministry of Economic Development and Innovation.

\end{document}